\documentclass{PoS}

\usepackage{placeins}
\usepackage{tabularx}
\usepackage{amsmath}
\usepackage{wrapfig}

\newcommand{\beq}{\begin{equation}}
\newcommand{\eeq}{\end{equation}}
\newcommand{\bea}{\begin{align}}
\newcommand{\eea}{\end{align}}
\newcommand{\beas}{\begin{align*}}
\newcommand{\eeas}{\end{align*}}

\newcommand{\bp}{{\bf p}}

\newcommand{\Tint}[1]{{\hbox{$\sum$}\!\!\!\!\!\!\!\int\,}_{\!\!\!\!\raise-0.9ex\hbox{$\scriptstyle{#1}$}}}

\title{What lattice QCD spectral functions can tell us about heavy quarkonium in the QGP}

\ShortTitle{Lattice QCD spectra of heavy quarkonium in the QGP}

\author{\speaker{Alexander Rothkopf}\\
        Institute for Theoretical Physics,  Heidelberg
  University, Philosophenweg 16, D-69120 Heidelberg, Germany\\
        E-mail: \email{rothkopf@thphys.uni-heidelberg.de}}      

\abstract{The bound states of a heavy quark and antiquark ($c\bar{c}, b\bar{b}$) are ideal probes to explore the quark-gluon plasma created in relativistic heavy-ion collisions at the RHIC and LHC. Not only have they become experimentally accessible with high precision but also efficient tools, so called effective field theories (EFT) have been developed to treat them theoretically. Here we present recent progress in understanding the in-medium behavior of heavy-quarkonium with the help of EFT's combined with non-perturbative and first principles simulations in lattice QCD. In particular we discuss computations of heavy quarkonium spectral functions with the help of Bayesian unfolding methods and the physics we can extract from them. Limitations and the underlying assumptions of the used approaches are pointed out. }

\FullConference{38th International Conference on High Energy Physics\\
		3-10 August 2016\\
		Chicago, USA}

\begin{document}


With run1 at the LHC concluded and run2 under-way experiment has provided an unprecedented amount of high precision data on heavy quark and antiquark bound states, so called heavy quarkonium ($c\bar{c}$ charmonium, $b\bar{b}$ bottomonium), in heavy-ion collisions. Two highlights \cite{Andronic:2015wma} in this regard at $\sqrt{s_{NN}}=2.76$TeV are the observation of relative excited states suppression for the bottomonium vector channel S-wave states by the CMS collaboration and the discovery by ALICE of a replenishment of yields for the vector channel ground state of charmonium $J/\psi$, compared to previous measurements at RHIC at $\sqrt{s_{NN}}=200$GeV. More recently there have been indications for a non-vanishing elliptic flow $v_2$ of $J/\psi$, which would hint at an at least partial kinetic equilibration of the quarkonium state with its surroundings.

The goal of theory is to contribute to an understanding of this rich phenomenology from first principles, which however requires idealization. In this contribution the underlying assumption is that of full kinetic equilibration of the heavy quarks, being embedded into a static thermal medium at a certain temperature. As this will be best justified for $c\bar{c}$ at low $p_T$ and mid rapidity in the following we focus on charmonium results only.

What makes heavy quarkonium amenable to theoretical treatment is the separation of scales between the heavy quark rest mass and the characteristic scales of its environment, such as temperature $T/m_Q\ll1$ and the characteristic scale of QCD $\Lambda_{\rm QCD}/m_Q\ll 1$. In the presence of such a separation ETF's allow one to simplify the language needed to describe the relevant physics at one of the lower scales, via the process of integrating out higher scales. In the case of quarkonium in the QGP e.g. the physics of heavy quark pair creation is irrelevant and thus one may go over to a non-relativistic language either of heavy-quark Pauli spinors in non-relativistic QCD (NRQCD) or that of heavy quarkonium wavefunctions (pNRQCD) \cite{Brambilla:2004jw}. The connection to the underlying QCD is made via the process of matching, where correlation functions with the same physical content in the EFT and in QCD are set equal at a given scale.

To evaluate field theories at the temperatures present in a realistic heavy ion collision $T\lesssim600$MeV, i.e.\ close to the chiral crossover $T_{PC}\approx155$MeV, perturbation theory cannot be used and one has to resort to numerical simulations, so called lattice QCD. One needs to keep in mind that real-time dynamics cannot be accessed directly in that approach, since simulations due to the notorious sign-problem are carried out in an artificial imaginary time direction.

To describe the in-medium dynamics of a quark-antiquark two-body systems in the appropriate language of quantum field theory one has to compute an in-medium spectral function $\rho(\omega)$. This quantity is related to the $r\to0$ limit of the Fourier transform of the so called forward heavy-quarkonium real-time correlation function $D^>(t,r)\equiv\langle \psi_{Q\bar{Q}}(t,r)\psi_{Q\bar{Q}}(0,r)\rangle$, i.e. the unequal time correlator of the $Q\bar{Q}$ wavefunction. A stable bound state appears in $\rho(\omega)$ as delta peak located at the mass of that particle, while resonances appear as Breit-Wigner type features with a finite width. At and above the energies at which it is possible to create open heavy-flavor mesons the spectrum contains a continuum. Important properties of an in-medium state can be learned from spectra, such as its in-medium binding energy, defined from the distance of the spectral peak to the open-heavy flavor threshold.

Here we report on recent progress in computing the in-medium spectral functions for charmonium using two different and complimentary approaches. Both face the same challenge that they need to extract a dynamical information from lattice QCD simulations via an unfolding procedure, which we attack using methods from Bayesian inference. Instead of the well known Maximum Entropy Method, we deploy here a more recent Bayesian reconstruction (BR) approach \cite{Burnier:2013nla}, specifically designed for the task of unfolding one-dimensional spectra.
 
 \section{Heavy quark potential from lattice QCD and in-medium spectra}

The first approach to quarkonium spectra is an indirect one and it uses the EFT pNRQCD. There one computes the in-medium potential between two heavy quarks at finite temperature and subsequently solves the Schr\"odinger equation for $D^>(t,r)$ and in turn for $\rho(\omega)$. While for a long time theorists resorted to purely real-valued model potentials, such as the color singlet free- or internal energies, it has become understood that neither of these quantities can represent the proper potential. The reason is that the matching procedure for pNRQCD has shown that the proper potential is related to the late time limit of a real-time QCD quantity, the rectangular Wilson loop
\beq
V(r)=\lim_{t\to\infty} \frac{i\partial_t W_\square(t,r)}{W_\square(t,r)}, \quad W_\square(t,r)=\Big\langle {\rm Tr} \Big( {\rm exp}\Big[-ig\int_\square dx^\mu A_\mu^aT^a\Big] \Big) \Big\rangle\label{Eq:VRealTimeDef},
\eeq
and its evaluation in resummed perturbation theory \cite{Laine:2006ns} revealed that it does take on complex values in the QGP at high temperatures. Eq.\eqref{Eq:VRealTimeDef} cannot directly be evaluated in lattice QCD but the technical concept of spectral function can help us to connect the Minkowski domain, in which the potential is defined, with the Euclidean domain in which the simulations are carried out \cite{Rothkopf:2011db} 
\begin{align}
\nonumber W_\square(\tau,r)\hspace{-0.1cm}=\hspace{-0.15cm}\int d\omega e^{-\omega \tau} \rho_\square(\omega,r)\,
&\hspace{-0.1cm}\leftrightarrow\hspace{-0.17cm}\, \int d\omega e^{-i\omega t} \rho_\square(\omega,r)\hspace{-0.1cm}=\hspace{-0.1cm}W_\square(t,r),\quad V(r)=\lim_{t\to\infty} \frac{ \int d\omega \omega e^{-i\omega t} \rho_\square(\omega,r)}{\int d\omega\, e^{-i\omega t} \rho_\square(\omega,r)}. \label{Eq:PotSpec}
\end{align}

\begin{figure}\vspace{-0.4cm}
\centering
\includegraphics[scale=0.3]{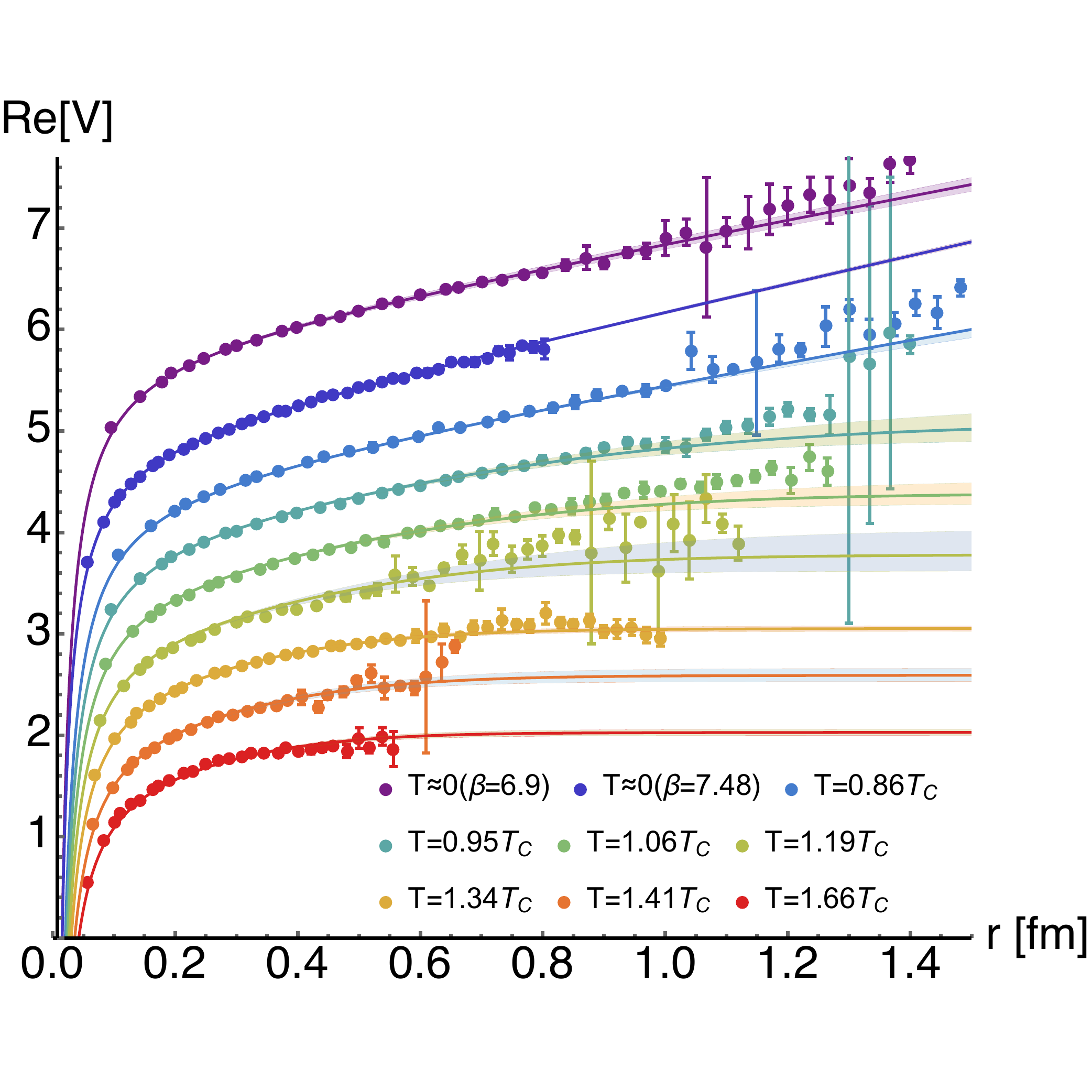}\hspace{0.5cm}
\includegraphics[scale=0.3]{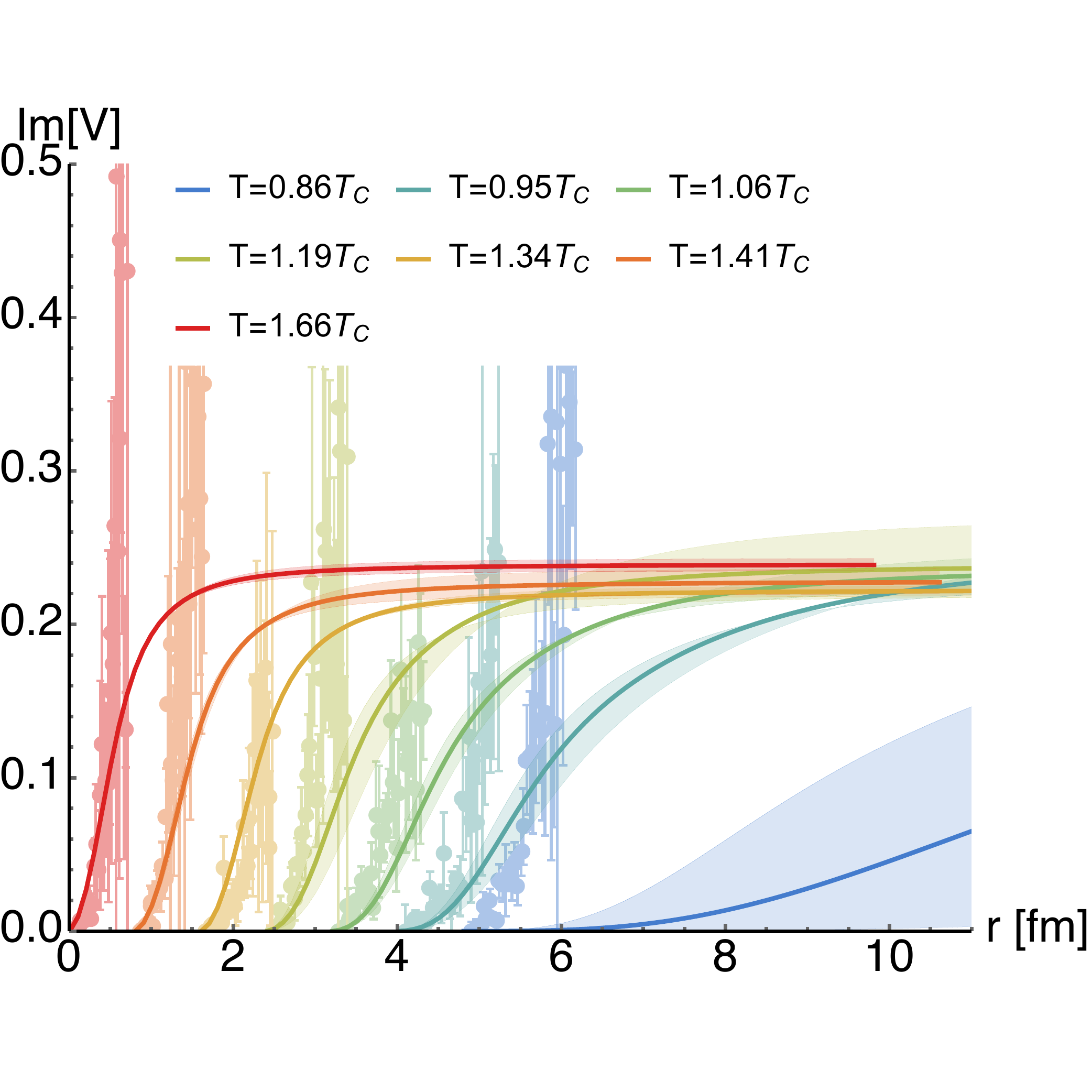}\vspace{-0.6cm}
\caption{ (left) Real- and (right) imaginary part of the proper static inter-quark potential at finite temperature from lattice QCD (solid points) with dynamical $u,d$ and $s$ quarks. Solid lines denote an analytic parametrization in which all temperature effects are encoded by a single T-dependent parameter $m_D$ the Debye mass.}\label{Fig:PropPot}
\end{figure}

It has been shown that the position and width of the lowest lying peak feature in the Wilson loop spectrum encode ${\rm Re}[V]$ and ${\rm Im}[V]$ respectively \cite{Burnier:2012az}. We have evaluated the potential in full QCD with dynamical $u,d$ and $s$ quarks (albeit with a relatively large $m_\pi=300$MeV) \cite{Burnier:2014ssa}, see Fig.\ref{Fig:PropPot}.

The gradual in-medium modification of ${\rm Re}[V]$ from a confining to a Debye screened behavior, as well as the appearance of an imaginary part $T\gtrsim T_C$ is visible. Using the concept of a generalized Gauss law we recently devised \cite{Burnier:2015nsa} an analytic parametrization of $V(r)$  (solid lines) that captures the in-medium behavior with a single temperature dependent parameter $m_D(T)$, the Debye mass.

\begin{wrapfigure}{r}{0.5\textwidth}\vspace{-0.5cm}
  \begin{center}
   \includegraphics[scale=0.4,trim=0 0 0 0.32cm, clip=true]{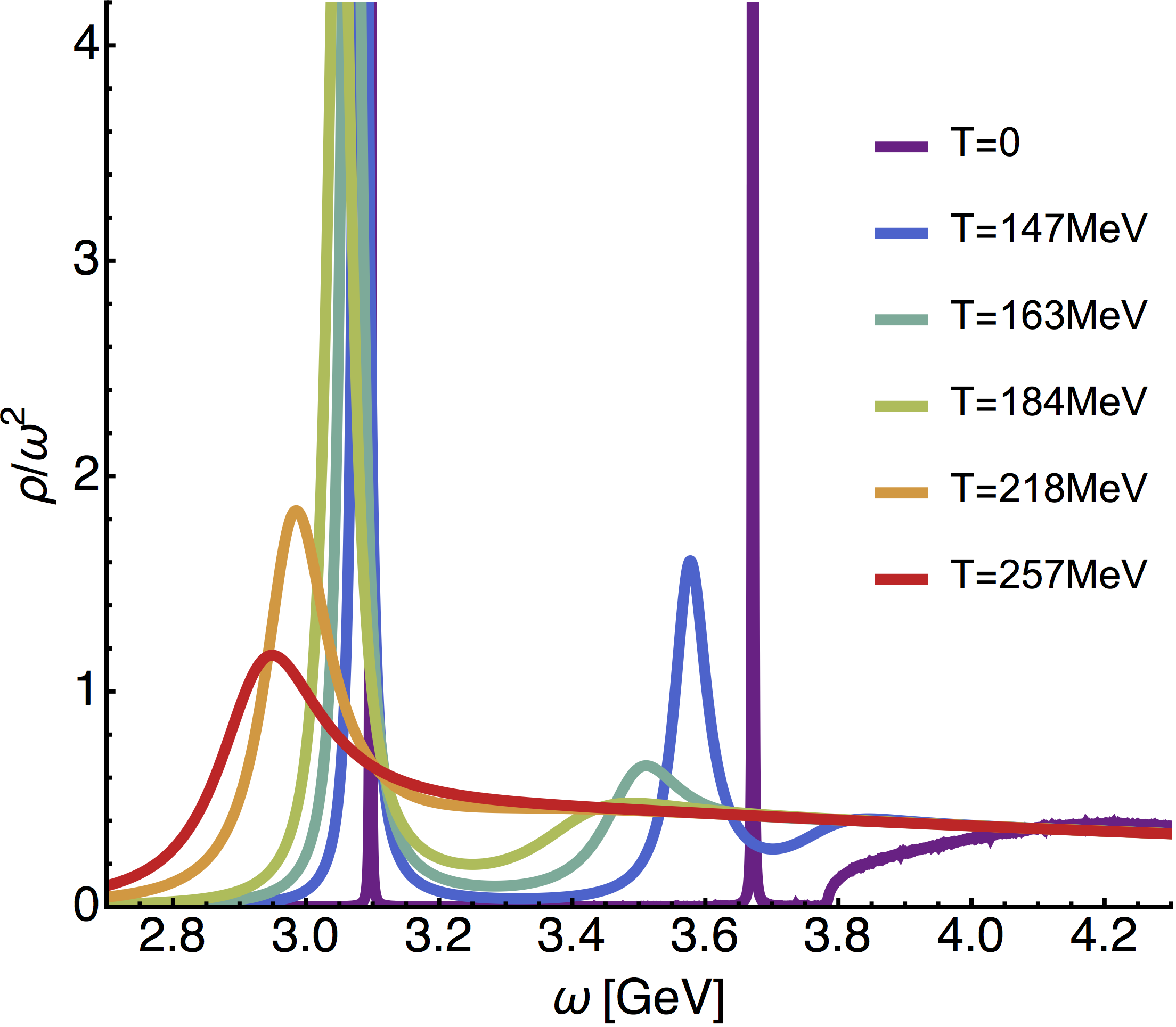}
  \end{center}\vspace{-0.8cm}
 \caption{Charmonium S-wave spectra based on the in-medium modification of the static inter-quark potential.}\label{Fig:CharmSpecs}\vspace{-0.2cm}
\end{wrapfigure}

Since we do not yet have continuum extrapolated lattice results for the potential we compute in the following the spectra \cite{Burnier:2015tda} using a phenomenological Cornell ansatz for the $T=0$ potential and implement on top the in-medium modification using $m_D$ obtained on the lattice. Using as mass for the charm quark $m_c=1.472$GeV appropriate for a computation in pNRQCD, we solve the Schr\"odinger equation for $D^>$ in Fourier space and evaluate $\rho(\omega)$ as shown in Fig.\ref{Fig:CharmSpecs}.
When increasing $T$, the former delta-peak like vacuum bound states $J/\psi$ and $\psi^\prime$ (dashed line) show a hierarchical modification pattern according to their vacuum binding energy. The features broaden and shift to lower masses. Their binding energy decreases, since the open-charm threshold also moves to lower masses as ${\rm Re}[V]$ weakens.

The concept of quarkonium melting that in studies based on real-valued model potentials appeared straight forward, becomes more involved once an ${\rm Im}[V]$ is present. I.e. there is no meaning in the concept of in-medium eigenstate and quarkonium dissolution must be understood as a inherently dynamical process. Using the well-defined spectral functions we may declare a state melted, e.g. once its width equals its in-medium binding energy leading e.g. to $T_{\rm melt}^{J/\psi}=213\pm13$MeV. In a simple wavefunction picture it would mean that after one oscillation the amplitude is damped by one factor of $e$. Note that even after melting so defined, a remnant of the former bound state may persist in the spectrum as threshold enhancement.

\begin{wrapfigure}{r}{0.5\textwidth}\vspace{-0.9cm}
  \begin{center}
   \includegraphics[scale=0.6]{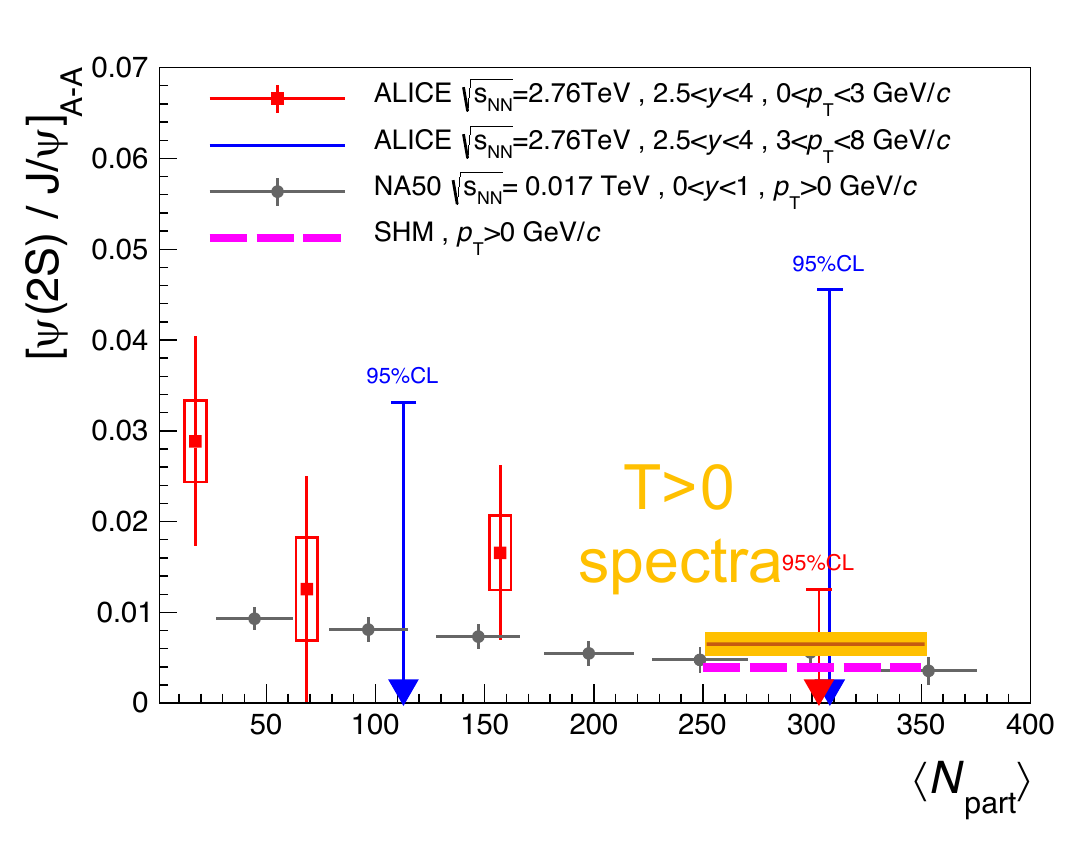}
  \end{center}\vspace{-0.9cm}
 \caption{Our $\psi'/J/\psi$ ratio (orange) as well as from the statistical model of hadronization (pink) and recent first measurements by ALICE (adapted from \cite{Andronic:2015wma}).}\label{Fig:CompFree}\vspace{-0.3cm}
\end{wrapfigure}$T_{\rm melt}$, while of theoretical interest, cannot be measured directly in experiment. Therefore we wish to go one step further and extract phenomenologically relevant information from the in-medium spectra. As many different models are able to reproduce e.g. the nuclear suppression factor $R_{AA}$ of $J/\psi$ it has been proposed to instead look at more discriminating observables, such as the $\psi^\prime/J/\psi$ ratio, which we estimate here in this fully thermal setting. Similar to the statistical model we assume an instantaneous freezeout scenario, where at $T_C$ the in-medium states abruptly go over into vacuum states. Therefore we wish to answer the question, how many vacuum states does the in-medium spectral peak at $T\approx 155$MeV correspond to? Our answer will be given in units of dilepton emission, which we can compute from the weighted area under the spectral peaks $R_{\ell\bar\ell}\propto \int dp_0 d^3\bp \frac{\rho(P)}{P^2}n_B(p_0)$. We then divide the in-medium rate of a state by the vacuum rate, establishing our estimate for the number of vacuum states produced. When carried out for both $\psi^\prime$ and $J/\psi$ we can form the ratio and obtain the value
\beq
\nonumber \left. \frac{N_{\Psi'}}{ N_{J/\Psi}} \right|_{T=T_C}=\frac{R_{\ell\bar\ell}^{\Psi'}}{ R_{\ell\bar\ell}^{J/\Psi}} \frac{ M_{\Psi'}^2 |\Phi_{J/\Psi}(0)|^2}{M_{J/\Psi}^2 |\Phi_{\Psi'}(0)|^2} =0.052\pm0.009, \,  \left.\frac{N_{\Psi'}}{ N_{J/\Psi}} \right|_{T=T_C} \frac{BR(\psi^\prime\to\mu^-\mu^+)}{BR(J/\psi\to\mu^-\mu^+)}=0.0069(11).
\eeq
A similar analysis for the production of P-wave states has been carried out recently in \cite{Burnier:2015tda}.

 \section{Direct spectra from lattice NRQCD}

As the previous computation of quarkonium spectra was based
\begin{wrapfigure}{r}{0.65\textwidth}\vspace{-0.5cm}
  \begin{center}
   \includegraphics[scale=0.6]{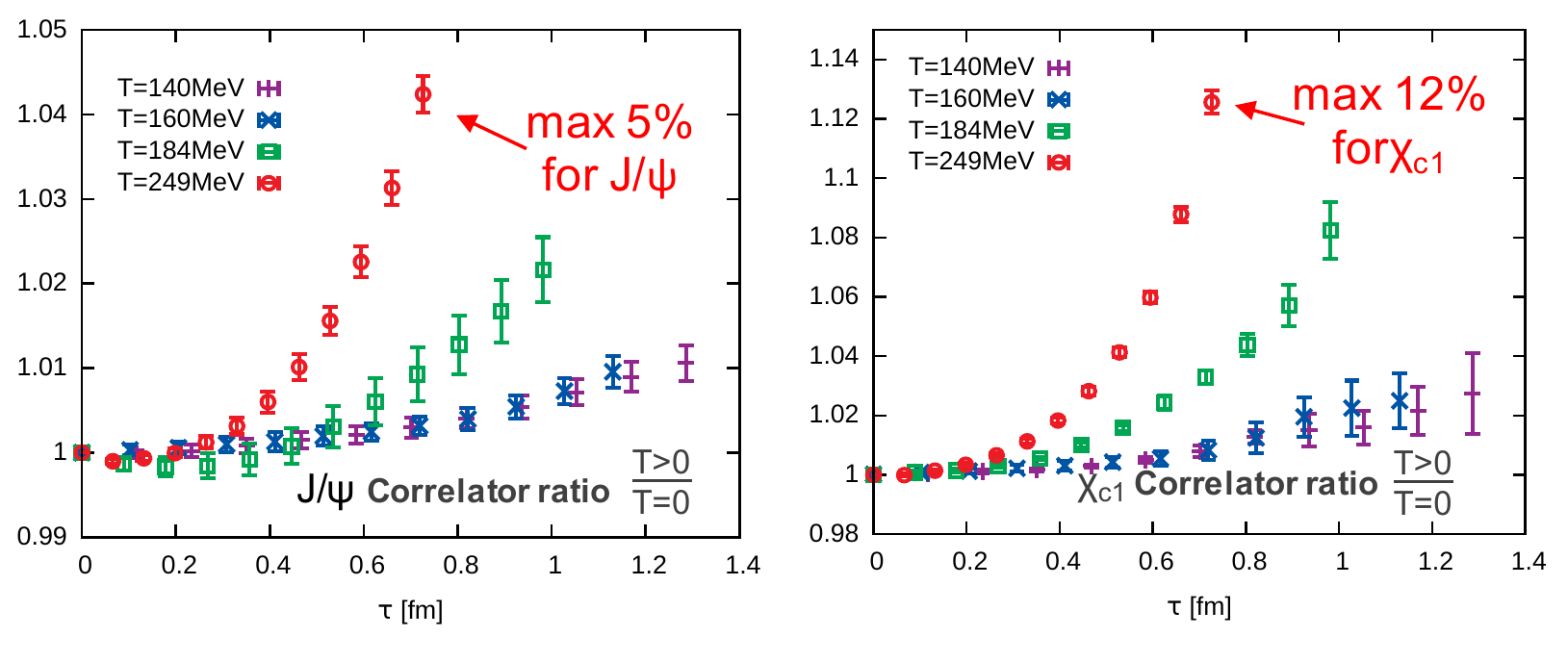}
  \end{center}\vspace{-1cm}
 \caption{Ratio of the $T>0$ and $T\approx0$ vector channel charmonium correlators in lattice NRQCD for the S-wave (left) and P-wave (right).}\label{Fig:CorrRatio}\vspace{-0.3cm}
\end{wrapfigure} 
 on the static in-medium potential which did not contain any finite-mass corrections it is paramount to crosscheck the results with an approach that does treat finite mass quarkonium directly. This is possible using lattice NRQCD, a well established technique used to also explore $T=0$ quarkonium (for details see \cite{Kim:2014iga}). It relies on a systematic expansion of the QCD action in powers of $1/m_Qa$, $a$ denotes the lattice spacing. In this approach the Euclidean current-current correlator $D(\tau)=\lim_{r\to0} D^>(-i\tau,r)$ can be simulated and the spectral functions has to be extracted via unfolding of $D(\tau)=\int_{-m_Q}^\infty d\omega\, e^{-\omega \tau}\rho(\omega)$.

\begin{wrapfigure}{r}{0.45\textwidth}\vspace{-0.9cm}
  \begin{center}
   \includegraphics[scale=0.5]{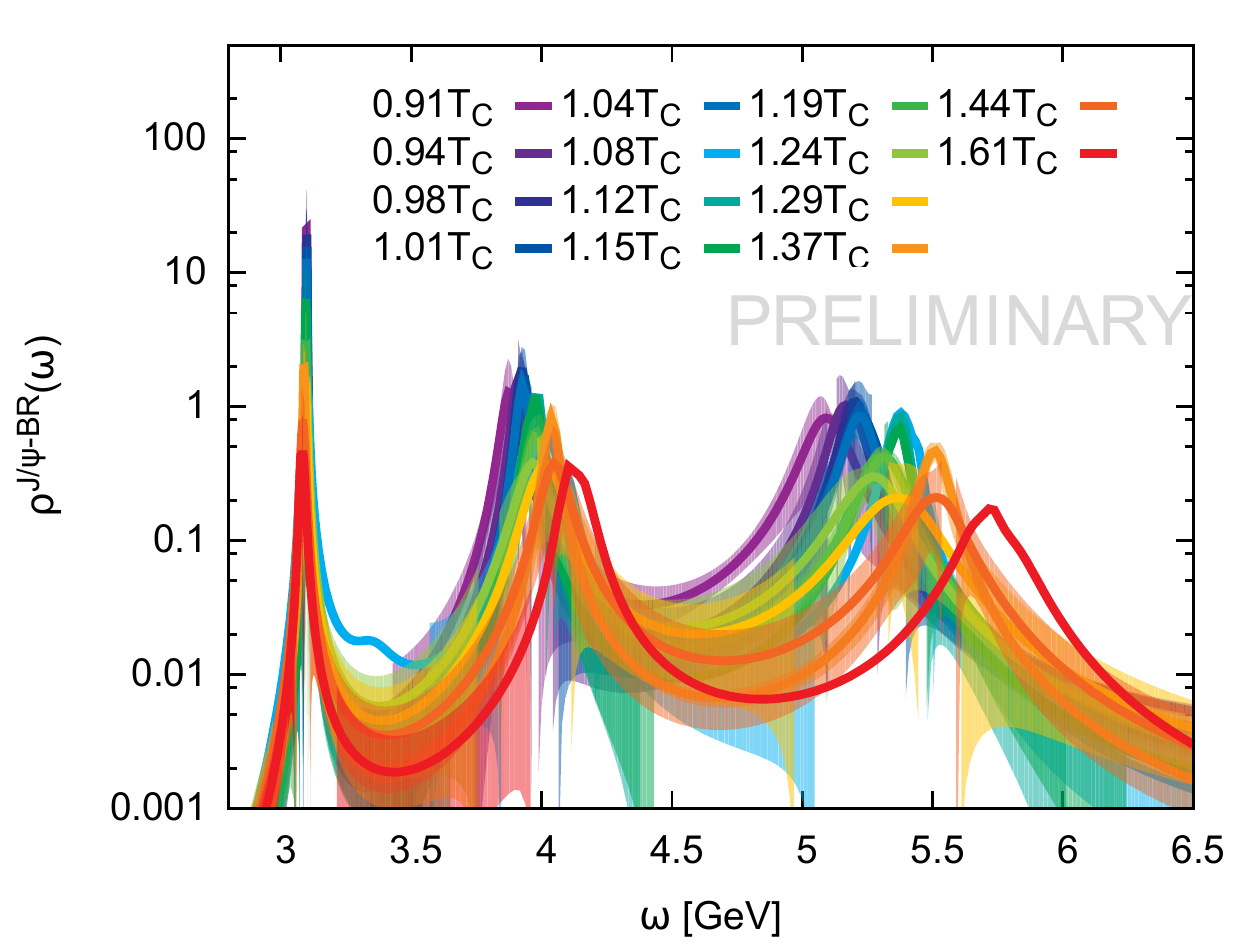}
  \end{center}\vspace{-0.8cm}
 \caption{Preliminary directly reconstructed spectrum of in-medium S-wave charmonium from lattice NRQCD.}\label{Fig:CharmSpecNRQCD}
\end{wrapfigure} 
Here we use realistic lattice simulations of finite temperature QCD with dynamical $u,d$ and $s$ quarks ($m_\pi=161$MeV), carried out by the HotQCD collaboration \cite{Kim:2014iga}. While the medium is captured very accurately the small lattice spacing requires the application of NRQCD to high order, we use $(1/m_Qa)^4$. Without performing a spectral reconstruction we can already learn about the in-medium modification of charm states, by inspecting the ratio of correlators $D^{T>0}(\tau)/D^{T\approx 0}(\tau)$ shown in Fig.\ref{Fig:CorrRatio}. If the ratio is unity no in-medium modification is present. Above $T_C^{\rm lat}=159$MeV the ratio shows a characteristic upward bend indicating in-medium effects. Note that the overall effect at $T=1.61T_C$ is small (5\% for $J\psi$,12\% for $\chi_c$) but appears to be ordered according to the vacuum binding energy of the ground state of a channel. 

A preliminary reconstruction of the S-wave channel is given in Fig.\ref{Fig:CharmSpecNRQCD}, where due to the small number of available simulation datapoints only the lowest lying feature may be considered robustly captured. One can see that qualitatively similar to the potential based study, with increasing temperature, the ground state gradually weakens and slightly shifts to lower frequencies. The quantitative determination of this in-medium modification however is still work in progress, in particular to determine the systematic uncertainties beyond the statistical errorbands shown.\vspace{-0.2cm}

\section{Summary}
The last decade has seen steady progress in both the experimental and theoretical exploration of in-medium heavy-quarkonium. On the one hand maturation of effective field theories have led to a systematic definition of the heavy-quark potential \cite{Burnier:2016mxc}, setting to rest the long lasting discussion about model potentials. On the other hand lattice QCD has arrived at realistic simulations of the QGP with light $u,d$ and $s$ quarks with almost physical pion mass. With progress in the extraction of real-time information from lattice simulations via Bayesian inference we are now able to much better leverage the non-perturbative predictive power of the lattice together with the intuitive language of EFT's to compute quarkonium in-medium spectra.

Here we reported on the recent extraction of the proper complex valued in-medium potential from full QCD and based on its in-medium modification we computed the charmonium spectra, observing characteristic hierarchical modification of states with respect to their vacuum binding energy. The peaks broaden and shift to lower mass as temperature rises. Declaring a state melted if its in-medium width equals its in-medium binding energy we are able to determine melting temperatures from the spectra. Furthermore by assuming an instantaneous freezeout scenario we estimated the $\psi^\prime/J/\psi$ ratio. As the potential currently does not contain finite mass corrections yet, we also wish to crosscheck the spectra from a direct determination in lattice NRQCD, where heavy but not static quarks are treated. While NRQCD correlators corroborate the picture of hierarchical in-medium modification, the relatively small number of simulated data points currently only allows us to capture the ground state peak feature with a quantitative determination of the in-medium mass and width being work in progress.

\end{document}